# Single-mode lasing from a single 7 nm thick monolayer of colloidal quantum wells in a monolithic microcavity

*Sina Foroutan-Barenji[1], Onur Erdem[1], Savas Delikanli[1], Huseyin Bilge Yagci[1], Negar Gheshlaghi[1], Yemliha Altintas[1,2], and Hilmi Volkan Demir[1,3]\**

[1]Department of Electrical and Electronics Engineering, Department of Physics, UNAM – Institute of Materials Science and Nanotechnology, Bilkent University, Ankara 06800, Turkey

[2]Department of Materials Science and Nanotechnology, Abdullah Gul University, Kayseri 38080, Turkey

[3]LUMINOUS! Centre of Excellence for Semiconductor Lighting and Displays, The Photonics Institute, School of Electrical and Electronic Engineering, School of Physical and Mathematical Sciences, School of Materials Science and Engineering, Nanyang Technological University, 50 Nanyang Avenue, Singapore 639798, Singapore

ABSTRACT. In this work, we report the first account of monolithically-fabricated vertical cavity surface emitting lasers (VCSELs) of densely-packed, orientation-controlled, atomically flat colloidal quantum wells (CQWs) using a self-assembly method and demonstrate single-mode lasing from a record thin colloidal gain medium with a film thickness of 7 nm under femtosecond optical excitation. We used specially engineered CQWs to demonstrate these hybrid CQW-VCSELs consisting of only a few layers to a single monolayer of CQWs and achieved the lasing from these thin gain media by thoroughly modeling and implementing a vertical cavity consisting of distributed Bragg reflectors with an additional dielectric layer for mode tuning. Accurate spectral and spatial alignment of the cavity mode with the CQW films was secured with the help of full electromagnetic computations. While overcoming the long-pending problem of limited electrical conductivity in thicker colloidal films, such ultra-thin colloidal gain media can help enabling fully electrically-driven colloidal lasers.

Lasers based on solution-processed materials have attracted intensive scientific and technological interest owing to their exceptional advantages including ease of fabrication, low-cost synthesis and processability and precise control of emission wavelength enabling lasing readily in different colors. As a result, a wide range of solution-processed materials of different classes have been studied for lasing, including organic semiconductors,[1] perovskites[2] and semiconductor nanocrystals.[3-5] Successful incorporation of such emitters into lasers thus far have helped expanding the use of lasers into new areas, such as medical and biological imaging,[6] chemical sensing,[7] lab-on-a-chip diagnostics,[8] and on-chip integrated photonic circuits.[9,10] Among solution-processable light-emitting materials studied for lasing applications, CQWs have gained considerable attention recently owing to their attractive properties including large absorption cross-section,[11] giant oscillator strength,[12] large material gain coefficient,[13] giant gain cross-section,[14] and ultra-narrow emission spectra stemming from pure carrier confinement in one dimension.[15] Several studies have been conducted to overcome fast Auger recombination in order to develop practical CQW lasers,[16-18] and type-II CQWs,[19] giant-shell CQWs,[17] and gradually interfaced core/shell CQWs,[18] have been introduced for this purpose. Auger recombination can also be in principle controlled by optimizing cavity designs and their interactions with the gain media. For example, coupling CQW layers to cavities with small mode volumes and high quality factors (Q-factors), such as micropillars, microdisks and photonic crystal cavities, can enhance the spontaneous emission due to the Purcell effect,[20] as a result providing means of overcoming fast Auger recombination. This issue requires careful design of cavities offering optimized optical mode overlap coefficients ($\Gamma$) and Q-factors.

Monolithic fabrication of ultra-thin cavity VCSEL provides opportunities for enhancing Q-factor and enabling single-mode lasing. Single-mode lasers purified from noise of other lasing modes are advantageous in practical applications and typically enable lower lasing thresholds compared to multimode lasers. In addition, monolithic CQW laser fabrication is crucial for enhancing the VCSEL performance because parasitic scattering in the cavity originating from multi-component setup is prevented along with the surface disruptions. Previously, CQW-VCSELs were fabricated by sandwiching drop-casted several-micrometer-thick CQW films between two mirrors resulting in bulky devices suffering from heat generation, exhibiting small Q factors,[5] and multi-mode lasing.[21] This further leads to increased surface roughness and formation of stacks, which in turn significantly increase the scattering losses within the cavity. For monolithic fabrication of CQW-VCSELs, deposition of uniform CQW films is required to tune the cavity thickness and the resonance wavelength as well as to obtain cavities with high Q-factors. Recently, our group has reported CQW film deposition with a self-assembly technique, which allows for thickness and orientation-controlled deposition of monolayers of CQWs on solid substrates with near unity surface coverage and sub-nm surface roughness across $cm^2$-large areas.[22,23] Our method is compatible with well-developed micro- and nano-fabrication technologies and can integrate CQWs with developed photonic components and in a dense footprint on a chip scale, which is critical for commercial mas-production. The uniform layering of top dielectric layers after CQW deposition and the monolithic chip-scale device fabrication needs the thermal stability of the underlying CQW film. CQW films with oleic acid ligands are appropriate for monolithic VCSEL fabrication because they are stable at temperatures up to 200 °C.[24]

Unfortunately, the poor confinement of the resonant mode in thin films of few monolayers could possibly lead to high losses, consequently making the lasing impossible. As a result, there are very few reports on lasing from thin films of colloidal nanocrystals. Roh *et al.*[25] achieved lasing with a threshold of 18.8 µJ cm$^{-2}$ from a quantum dot (QD) gain medium as thin as 50 nm using a distributed feedback (DFB) resonator. Kim *et al.*[26] reported single-mode lasing from a colloidal QD gain medium only 35 nm thick with a threshold of 20 mJ cm$^{-2}$. Achieving lasing from films of CQWs thinner than hundreds of nanometers,[18,27] which has the advantage of giant modal gain compared to all other types of nanocrystals (QDs, nanorods and nanowires),[28] has not been previously accomplished.

In this study, we demonstrate the single mode lasing by using femtosecond optical pumping pulses from precisely engineered monolithic VCSELs having dense ultrathin films of CQW monolayers obtained by self-assembly technique. Here, mode tuning films present in the monolithic VCSELs were designed to achieve high Q-factors as well as accurate spectral matching with the gain spectrum with the help of full electromagnetic simulations. Owing to the uniformity and thermal stability of our ultra-thin CQW films used in these devices, the top DBRs could be deposited at temperatures above 150 °C, which ensures the fabrication of high Q-factor VCSELs. We observed a lasing threshold of 112 µJ cm$^{-2}$ owing to the resulting high Q-factor of ~760 for a device having only 7-nm-thick CQW layer, which is the thinnest lasing demonstration among all colloidal nanocrystals reported to date to the best of our knowledge. We were also able to deposit multiple CQW monolayers uniformly in a cavity using self-assembly method and demonstrated lasing with a lower threshold of 78 μJ cm$^{-2}$ from a VCSEL with four CQW monolayers having total thickness of ~28 nm. It is worth noting that the CQW films in our devices were as thin as those used for electrical operation of colloidal light-emitting diodes.[29,30] Further cavity optimization in addition to additional improvement in CQW performance may enable the realization of electrically-driven CQW lasers, which is a long-term goal.

**Results and Discussion.** CdSe/Cd$_{0.34}$Zn$_{0.66}$S core/giant alloyed hot-injection (HI) shell CQWs have been synthesized according to our recent recipe based on well-adjusted precursor and ligand concentrations.[17,24] To produce highly emissive, spectrally narrow and environmentally stable near-unity efficiency core/shell CQWs, we have grown giant graded-shell on seed CdSe CQWs by the HI technique.[31-34] These novel HI-grown graded-shell CQWs were shown to exhibit near-unity photoluminescence quantum yield (PL-QY) and high photo- and thermal-stability,[24] and the device performance of these materials that have been recently reported with record high external quantum efficiency in light-emitting diodes[32] and high optical gain and ultralow lasing thresholds.[17,35] **Figure 1**a presents PL and absorption spectra of our CQWs. The inset of Figure 1a shows a representative transmission electron microscopy (TEM) image of these CQWs. The PL-QY of our CQWs used in this work was measured to be 97%. The two excitonic features in the absorbance spectra located at 632 and 574 nm correspond to the electron-heavy hole and electron-light hole transitions, respectively. The PL emission having a full-width at half-maximum (FWHM) of 25 nm peaks at 640 nm. The TEM image reveals that these CQWs have the average dimensions of ($l$

× w × h) of 17.6 ± 1.6 × 16.9 ± 1.2 × 4.6 ± 0.5 nm$^3$, and their total thickness is 7 nm if the thickness of the capping ligands are taken into account.

Time-resolved PL (TRPL) data of the synthesized CdSe/Cd$_{0.34}$Zn$_{0.66}$S core/giant alloyed HI-shell CQWs are given in Figure 1b. The TRPL data were fitted with bi-exponential decay function and the resulting amplitude average lifetime is 24 ns. The lifetime of the core-only sample is only 1.8 ns. The dangling bonds and defects on the core surface are fully passivated by our shell growth strategy. Along with the increase in the lifetime, the PL-QY of the core-only samples increases from 20% to 97% while growing the shell on it. This increase in lifetime accompanied with enhancement in the PLQY suggests passivation of the trap states present in the core by the high quality shell deposition. These findings are also supported by the prolonged biexciton lifetime exceeding 1.2 ns and greatly suppressed Auger recombination that is one of the significant bottleneck to overcome for the population inversion in practical laser applications.[17,36] Figure 1c shows X-ray photoelectron spectra (XPS) of the CQWs to clarify chemical structure of our core/shell sample by providing elemental percentage of each of its individual elements with their carefully fitted and identified spin-orbit components. The atomic percentages of the Cd and Zn in the shell are found to be 34 and 66, respectively, from the calculation procedure (see the Supporting Information).

The optical gain properties of these CQWs were studied in our previous work, in which we showed the capability of amplified spontaneous emission (ASE) from bi-CQW layers with a thickness of only 14 nm.[23] This observation shows the possibility of lasing action from ultra-thin CQW gain media provided that coupled into an effective cavity. In this work, we fabricated CQW-VCSELs with ultra-thin CQW gain media and investigated their lasing characteristics. Here each device consists of a bottom distributed Breagg reflector (DBR) constructed with alternating layers of SiO$_2$ and TiO$_2$, a cavity formed by single or multiple layers of CQWs, a mode-tuning dielectric layer, and a top DBR containing SiO$_2$ and TiO$_2$ layers. According to the Bragg–Snell law at normal incidence, the thickness of SiO$_2$ ($t_l$) and TiO$_2$ ($t_h$) layers were chosen as quarter-wavelength of the emission peak of CQWs ($\lambda_{max}$ = 640 nm), which are equal to 106 and 64 nm, respectively. The effective length accounting for an almost complete decay (e$^{-2}$= 0.14) of the field intensity in a planar VCSEL structure can be calculated using

$$L_{eff} = L_{MC} + 4(t_l + t_h)\frac{n_{eff}}{(n_h - n_l)} \qquad (1)$$

where $L_{MC}$ is the geometrical thickness of the microcavity, $n_l$ is the refractive index of low-index material, $n_h$ is the refractive index of high-index material (see the Supporting Information for measured refractive indices), and $n_{eff}$ is the effective refractive index of the DBRs.[37] The effective index $n_{eff}$ can be obtained by

$$n_{eff} = \sqrt{\frac{t_l}{t_l + t_h}n_l^2 + \frac{t_h}{t_l + t_h}n_h^2} \qquad (2)$$

taking $L_{MC}$ = 7 nm for a cavity with a monolayer of CQWs sandwiched between DBRs with TiO$_2$/SiO$_2$ bi-layers results in $L_{eff}$= 1.3 µm. The transmittance of 11 bilayers DBRs with a length of 1.87 µm were measured to reach ~0.1% at $\lambda_{max}$ (Figure S1), verifying the high reflectance of the fabricated mirrors, which is required for achieving high cavity Q-factor, defined as Q=$\lambda$/FWHM. Figure S1 also shows that the transmittance of DBR is ~90% at the excitation wavelength (400 nm), which is needed for efficient pumping of the gain medium, too.

To achieve lasing in the cavity, the cavity resonant mode $\lambda_{res}$ should be spectrally aligned with the peak of CQW gain $\lambda_{max}$. Also, the optical field overlap $\Gamma$ with the gain medium should be maximized in order to achieve sufficient net modal gain given by $g_{modal} = \Gamma g_{material} - \alpha$, where

$g_{material}$ is the material gain and $\alpha$ is the total modal loss. The Γ-factor, and Q-factor as well as spectral behavior of the cavity were studied via numerical simulations based on finite-difference time-domain (FDTD) method (see the Methods section). All device parameters were carefully analyzed and optimized prior to the device fabrication for achieving the best possible performance in our numerical calculations. As shown in **Figure 2**a, $\lambda_{res}$ can be modulated continuously from ~600 to ~680 nm by increasing the thicknesses of the CQW films from a single monolayer up to 10 monolayers. This is the result of phase shift imparted by increasing of the cavity thickness. Adding a monolayer of CQWs with a thickness of $\Delta t$ = 7 nm to the cavity redshifts the resonance wavelength by a relatively large amount of $\Delta\lambda_{res}$ = +13 nm ($\Delta\nu_{res}$ = -11 THz) at $\nu_{res}$ = 493 THz. The shift $\Delta\nu_{res} = - \nu_{res} \Delta t / t_{eff}$ is proportional to effective cavity length $t_{eff}$, which is found to be ~ 315 nm (Figure 2e). The Γ factor increases by the thickness of CQW film, however the shift of the cavity mode to the edge of DBR stop-band reduces the reflection of the DBR and subsequently the Q-factor of the cavity. For our DBR design, the Γ×Q product reaches to a peak value at about 640 nm and reduces when the $\lambda_{res}$ is further increased by increasing the CQW film thickness.

Considering the ultrathin cavity thickness and resulting large free spectral range, $\lambda_{res}$ needs to be tuned to reside in the FWHM of CQW emission. To work around this problem, additional dielectric layers were incorporated between the mirrors, which provides us with the ability to tune $\lambda_{res}$ while maintaining a high Γ×Q product at the desired wavelength. We studied the effect of adding an extra layer of $SiO_2$, $Al_2O_3$ and $TiO_2$ inside the cavity containing different numbers of CQW layers. As shown in Figure 2b (top panel), Γ factor decreases as the thickness and refractive index of the mode-tuning layer increases. Increasing the thickness broadens the mode spatially and increasing the refractive index localizes the mode peak away from CQW film, and hence, reduces the effective mode overlap. The peak of Q-factor is related to the refractive index of the material, which happens at thinner films for the higher index. Figure 2c shows the calculated Γ×Q product for a cavity with a monolayer of CQWs using different mode-tuning materials. These calculations reveal that $Al_2O_3$ provides the best performance and the highest Γ×Q.

Based on our computational results, we chose to deposit 30 nm of $Al_2O_3$ on top of a monolayer of CQWs to match $\lambda_{res}$ with the peak emission of our CQWs. The simulated Q-factor of ~9,000 in this case shows a great potential for coupling the emission from CQW layer to the nanocavity, which can enhance the spontaneous emission due to the Purcell effect. Figure 2d shows the electric field intensity of the resonant mode at 639 nm for the designed monolayer CQW cavity. Figure 2e shows one period of this mode with a length of 340 nm, which indicates an effective mode index of 1.88.

We fabricated four sets of VCSELs having one to four CQW monolayers in the cavity. To tune the resonant mode of each VCSEL within the FWHM of the CQW emission, an extra layer of $Al_2O_3$, pf which the thickness is optimized according to the number of CQW layers, was deposited. The CQW monolayers were deposited through our liquid−air interface CQW self-assembly method, which offers us the capability of depositing one CQW layer at a time with near-unity surface coverage and sub-nm surface roughness.[22] The mode-tuning $Al_2O_3$ layer was laid down through atomic layer deposition at 150 °C. As the CQWs used in this study have oleic acid ligand and are thermally stable up to 200 °C, maintaining their PL,[24] it is possible to use the high-temperature deposition techniques on top of the CQWs to obtain uniform film quality in top layers.

**Figure 3**a shows the cross-sectional TEM image of one such VCSEL (top panel) and a magnified image of the cavity consisting of one monolayer of CQWs and $Al_2O_3$ mode-tuning layer (bottom panel). As shown in this image, the film in the fabricated device is highly uniform resulting in high cavity Q-factors. The scanning electron microscopy (SEM) image of a monolayer of self-

assembled CQWs is shown in Figure 3b, which exhibits near-unity surface coverage. Figure 3c depicts the simulated transmission spectrum for the four fabricated CQW-VCSELs. According to our simulations, a cavity mode exists within a wide range of photonic band gap in our designed structure. Hence, the cavity modes of all devices were successfully tuned to reside within the gain band of the CQWs. For a cavity with a monolayer of CQWs, adding 30 nm thick $Al_2O_3$ film causes a red shift in the mode from 595 to 639 nm. Similarly, for a cavity with 2 monolayers of CQWs, adding 30 nm of $Al_2O_3$ red-shifts the mode from 608 to 649 nm. With the addition of 10 nm of $Al_2O_3$, the mode in the cavity having three monolayers of CQWs red-shifts from 620 to 638 nm and, similarly, the mode red-shifts from 632 to 647 nm in the case of 4 monolayers of CQWs.

Figure 3d shows the measured PL spectra of the fabricated CQW-VCSELs at a pump fluence of ~119 µJ cm$^{-2}$ together with the simulated results. The experimentally observed lasing peak display an excellent match with our expectation from the numerical simulations. To find out the lasing threshold in each VCSEL, the cavities were excited with a pulsed laser at 400 nm with a pulse width of ~110 fs and a repetition rate of 1 kHz, by varying the pump fluence (see the methods section). The measured spectra from CQW-VCSELs having one to four monolayers of CQWs are presented in **Figures 4**a-d. Because of the ultra-thin cavity and high Q-factor in all cases, a single longitudinal mode exists in all devices. The linewidths extracted from Gaussian fits for the measured lasing spectra are plotted as a function of the pump fluence in Figures 4e-h (bottom panel). The onset of lasing is marked by a sharp increase in the PL intensity at the lasing threshold. By fitting the data with a linear function, we obtained a threshold pump energy density of ~112 µJ cm$^{-2}$ for the VCSEL having a monolayer of CQW layer (VCSEL1). The lasing threshold gradually decreases with the increasing number of CQW monolayers and reduces to 78 µJ cm$^{-2}$ for the VCSEL4 having four monolayers of CQWs. The cavity linewidth of VCSEL1 narrows down to 0.8 nm at the lasing threshold and reaches a minimum value of 0.5 nm at the fluence of 114 µJ cm$^{-2}$. Then, the linewidth gradually increases as we further keep increasing the pump fluence. For example, the linewidth widens to ~0.6 nm at the pump fluence of 119 µJ cm$^{-2}$. This broadening at high pump fluences can be attributed to the temperature fluctuation and/or refractive index modulation due to nonlinear effects under hard pumping, which may result in the cavity mode drift near the cavity resonance wavelength. Similar behavior has been observed for VCSEL2-4. The peak wavelengths extracted from Gaussian fits for the measured lasing spectra is shown in Figures 4e-h (bottom panel). For all VCSELs the peak wavelength deviation with increasing pump intensity is less than 1 nm, which shows excellent stability in emission wavelength. The small amount of deviation can be resulting from the refractive index modulation due to nonlinear effects, heating effects and noise. The lasing characterizations of the VCSELs are listed in Table 1 for all CQW film thicknesses studied.

The observed reduction in the lasing threshold with the increasing thickness of gain medium stems from the increased resonant mode overlap with the gain medium. Furthermore, in each of our VCSELs, most of the pump beam (more than 90%) was not absorbed by the ultra-thin CQW film, but rather was transmitted through or reflected off the device, which would also require higher pumping energy. By simply increasing the CQW film thickness, in addition to Γ-factor enhancement, the pump beam absorption in the gain medium could be increased, thereby increasing the modal gain at the same pump power. All CQW-VCSELs exhibited relatively low lasing thresholds on the order of tens of µJ cm$^{-2}$ which can be attributed to the spectral matching of the cavity mode and the maximized Γ- and Q-factors. Furthermore, a key factor for this high performance is the exploitation of hot-injection shell (HIS) growth of CQWs, which simultaneously enables a near-unity photoluminescence quantum yield (PLQY), reduces nonradiative channels, ensures smooth films, and enhances the stability.

**Conclusion**

In this study, we fabricated monolithic CQW-VCSELs and demonstrated single-mode lasing using densely-packed self-assembled films of CdSe/Cd$_{0.34}$Zn$_{0.66}$S core/giant alloyed HI-shelled CQWs as thin as 4.6 nm (if 2.4 nm organic ligands are not taken into account). Microcavities with the highest possible Q-factors as well as accurate spectral matching with the gain spectra were designed by employing numerical modeling. The high packing density of smooth films of the self-assembled CQWs, the high Q-factor of the cavity and the superior optical properties of core/shell CQWs synthesized with our hot injection technique having near-unity PL quantum yield enable low threshold single-mode lasing from ultra-thin films of these CQWs. The demonstrated CQW films used in this work are sufficiently thin to possibly enable electrical carrier injection, given that they are combined with proper carrier-transporting layers. The findings of this work may pave the way for realization of ultra-thin electrically-driven colloidal laser devices, providing critical advantages including single-mode lasing and high electrical conduction.

ASSOCIATED CONTENT

**Supporting Information**. Synthesis of CdSe/CdZnS CQWs, ellipometry measurement results, full electromagnetic simulation of CQW-VCSELs, details of monolithic fabrication of CQW-VCSEL, lasing measurement method, polarization characteristics of VCSEL.

AUTHOR INFORMATION

**Corresponding Author**

*E-mail: volkan@bilkent.edu.tr, hvdemir@ntu.edu.sg, volkan@alumni.stanford.org

Notes

The authors declare no competing financial interest.

ACKNOWLEDGMENT

The authors acknowledge the financial support from the Singapore National Research Foundation under the program NRF-NRFI2016-08 and in part from TUBITAK 115E679. The authors thank Mr. Mustafa Guler for his assistance in TEM imaging of the as-synthesized NPLs and preparation of the TEM cross-sectional sample and Dr. Gokce Celik for her help on the ellipsometric measurements. O.E. acknowledges TUBITAK for the financial support through BIDEB 2211 program. H.V.D. gratefully acknowledges support from TÜBA.

**Table 1.** Lasing parameters of our CQW-VCSELs.

| VCSEL | Number of CQW layers | CQW film thickness (nm) | $Al_2O_3$ film thickness (nm) | Lasing threshold, $\phi_{th}$ (µJ cm$^{-2}$) | Lasing wavelength, $\lambda_{res}$ (nm) |
|---|---|---|---|---|---|
| 1 | 1 | 7 | 30 | 112 | 640 |
| 2 | 2 | 14 | 30 | 107 | 649 |
| 3 | 3 | 21 | 10 | 102 | 638 |
| 4 | 4 | 28 | 10 | 78 | 648 |

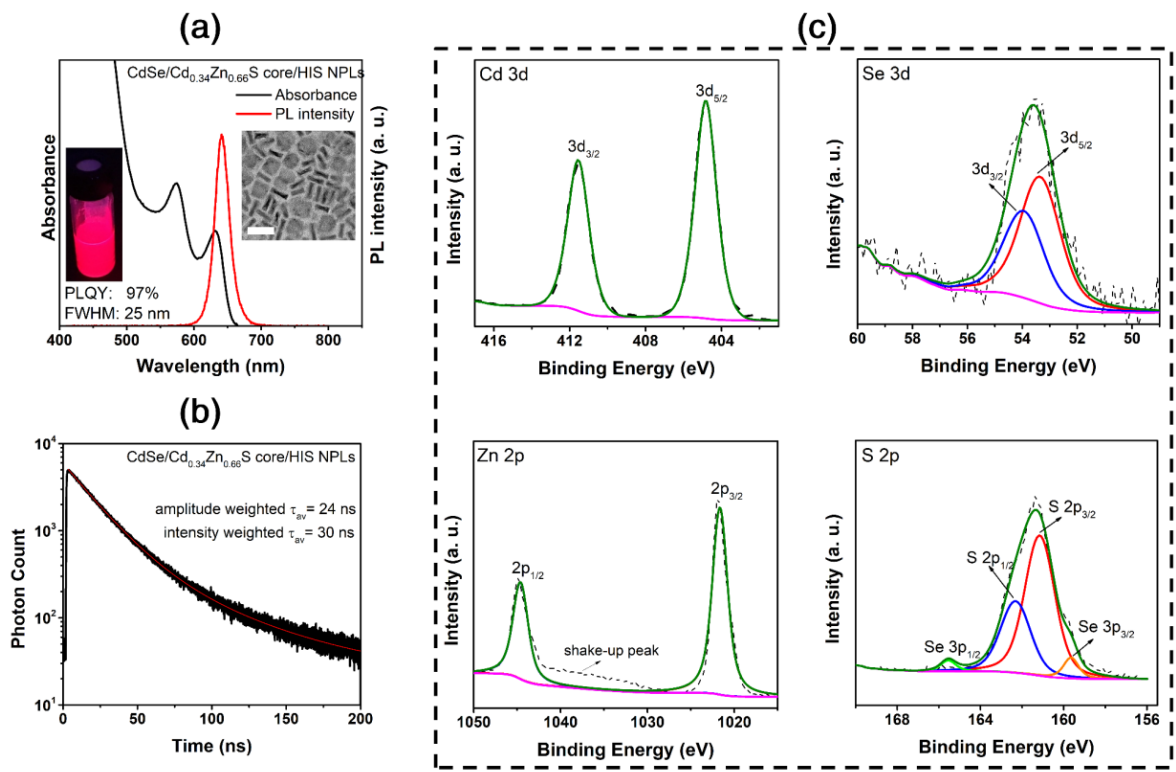

Figure 1. (a) Photoluminescence (PL) and absorption spectra of the CdSe/Cd$_{0.34}$Zn$_{0.66}$S core/giant alloyed HI shell-grown CQWs. Inset: High-resolution transmission electron micrograph of these CQWs (scale bar: 30 nm) (b) time-resolved PL spectra (in solution), (c) X-ray photoelectron spectra (XPS) of CdSe/Cd$_{0.34}$Zn$_{0.66}$S core/giant alloyed HI shell CQWs.

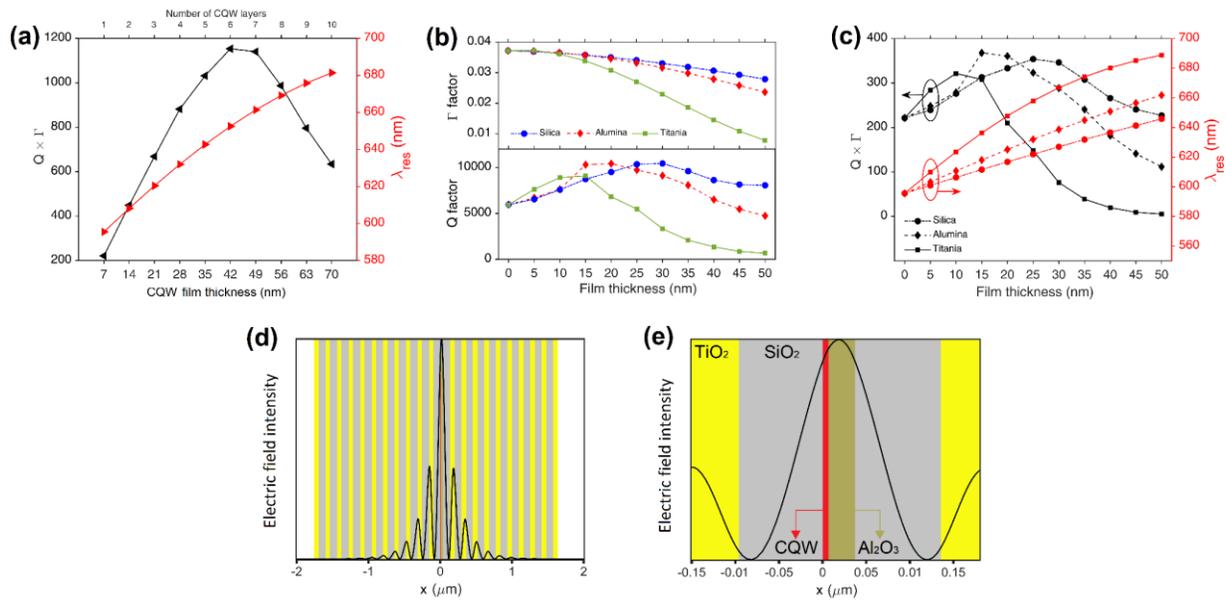

Figure 2. Product of Q-factor and Γ-factor and the corresponding resonant wavelength of our cavity (a) as a function of the CQW film thickness, (b) Q-factor and Γ-factor of the cavity separately for a monolayer of CQWs as a function of the mode tuning film (silica, alumina, and titania) thickness. (c) The product of Q-factor and Γ-factor and the corresponding resonant wavelength of the cavity for a monolayer of CQWs as a function of the mode-tuning film (silica, alumina, and titania) thickness. (d) Electric field intensity profile of the resonant mode inside a cavity with a monolayer of CQWs and 30 nm of alumina. (e) Magnified plot of main lobe of the electric field intensity inside the cavity covering CQW and Al2O3 layers at the center shown in Figure 2d.

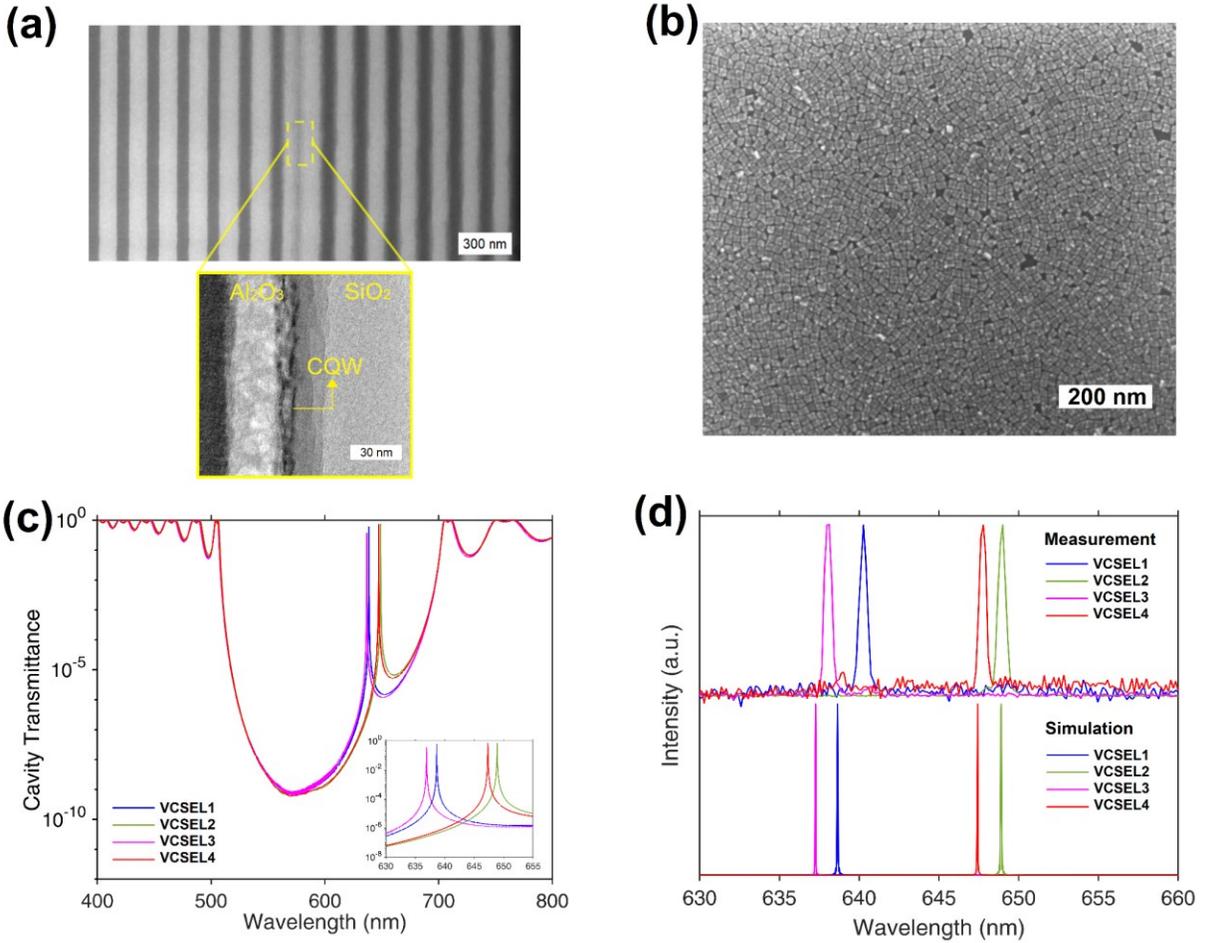

Figure 3. (a) Cross-section TEM image of CQW-VCSEL (top panel) and magnified image of a cavity (bottom panel) with a 7 nm thick (a monolayer-thick) CQW layer. (b) Top-view SEM image of a self-assembled CQW monolayer. (c) Simulated cavity transmission for CQW-VCSEL1, 2, 3, and 4, which possesses 7, 14, 21, and 28 nm thick CQW layers and exhibit cavity modes at 639, 649, 637, and 647 nm, respectively. The inset shows the magnified range containing the resonant modes. (d) Spectra measured at a pumping energy density of 119 µJ cm$^{-2}$ for CQW VCSELs (left panel) and design expectations from FDTD simulations (right panel).

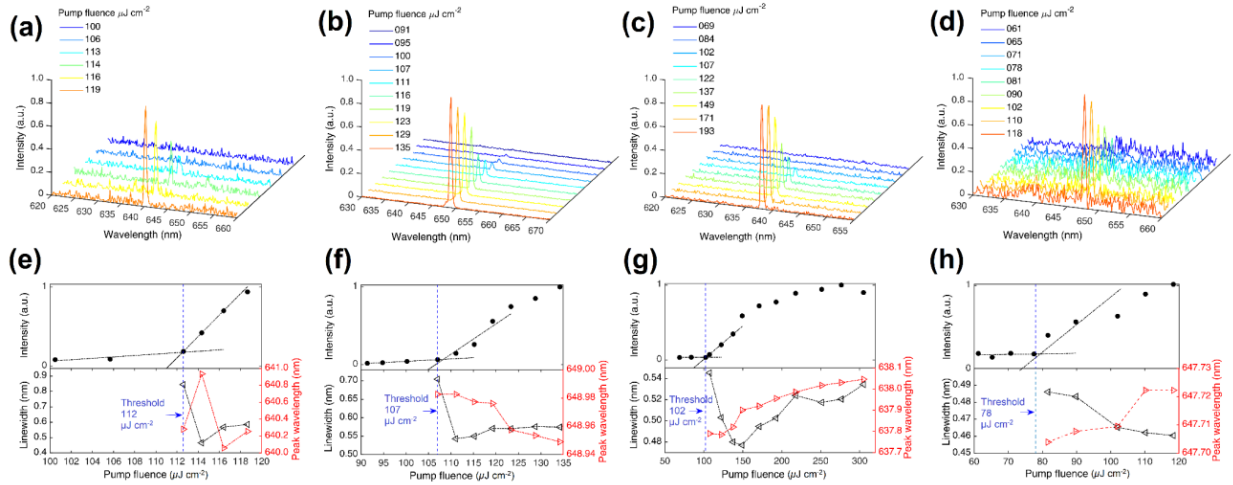

Figure 4. Measured spectra of CQW-VCSELs at various pumping energy densities. (a) VCSEL1, (b) VCSEL2, (c) VCSEL3, and (d) VCSEL4. Peak emission intensity (top panels), linewidth and peak emission wavelength (bottom panels) extracted from Gaussian fits for the lasing spectra, as a function of the pump fluence for (e) VCSEL1, (f) VCSEL2, (g) VCSEL3, and (h) VCSEL4.